\documentclass[reprint,superscriptaddress,amsmath,amssymb,aps,prl]{revtex4-2}

\usepackage{graphicx}

\usepackage{color}
\usepackage{tensor}
\usepackage[all,cmtip]{xy}
\usepackage{dcolumn}
\usepackage{bm}

\usepackage{dsfont}
\usepackage{enumitem}

\makeatletter
\newcommand*\bigcdot{\mathpalette\bigcdot@{.5}}
\newcommand*\bigcdot@[2]{\mathbin{\vcenter{\hbox{\scalebox{#2}{$\m@th#1\bullet$}}}}}
\makeatother

\usepackage{tensor}
\usepackage{physics}
\usepackage{fontawesome}
\usepackage{slashed}
\usepackage{amsmath,amssymb,calc, amsthm,bbm, epsfig,psfrag, mathtools}
\usepackage[normalem]{ulem} 

\usepackage{latexsym,bm,amsfonts}
\usepackage{graphicx, enumerate}
 \usepackage[all,cmtip]{xy}

\newcommand{\arXivlink}[1]{\href{http://arXiv.org/abs/#1}{arXiv:#1}}

\allowdisplaybreaks

\newcommand{\Comment}[1]{{}}
\definecolor{darkblue}{rgb}{0.15,0.35,0.55}
\definecolor{reddish}{rgb}{0.65, 0.2, 0.2}
\usepackage[linktocpage=true]{hyperref}
\hypersetup{
colorlinks=true,
citecolor=darkblue,
linkcolor=reddish,
urlcolor=darkblue,
pdfauthor={},
pdftitle={},
pdfsubject={}
}

\def\be{\begin{equation}}
\def\ee{\end{equation}}
\def\bea{\begin{eqnarray}}
\def\eea{\end{eqnarray}}

\newfont{\goth}{ygoth.tfm scaled 1200}                   

\def\1{{(1)}}
\def\2{{(2)}}
\def\3{{(3)}}

\newcommand{\alg}[1]{\mathfrak{#1}}

\newcommand{\ba}{\begin{array}}
\newcommand{\ea}{\end{array}}

\def\double #1{#1{\hbox{\kern-2pt $#1$}}}

\newcommand{\bsubeq}{\begin{subequations}}
\newcommand{\esubeq}{\end{subequations}}

\begin{document}


\title{
{\boldmath{$AdS_3$ integrability, Sine-Gordon and fractional supersymmetry}}}

\author{Alessandro Torrielli}
\email{a.torrielli@surrey.ac.uk}
\affiliation{School of Mathematics and Physics, University of Surrey, Guildford, GU2 7XH, UK}%

\date{\today}

\begin{abstract}
The massless $S$-matrix of the pure RR $AdS_3\times S^3 \times T^4$ theory is very similar but also crucially distinct from Fendley-Intriligator's famous ${\cal{N}}=2$ $S$-matrix, in turn related to Sine-Gordon taken at a special coupling. In this article we review the reason why supersymmetry emerges but with a fractional statistics of the particles. We then use this to obtain an interpolating $S$-matrix between massless $AdS_3$ and Sine-Gordon, and further find two interpolating $S$-matrices between these and the mixed-flux relativistic $S$-matrix, solving the Yang-Baxter equation all the way in-between. They are all relativistic invariant and braiding unitary, can be written in terms of particles with fractional statistics, and embed all the $AdS_3$ models into a parent integrable system. This also achieves the reformulation of all the $AdS_3$ integrable scattering problems in the domain of fractional statistics.\footnote{\ttfamily{DMUS-MP-25/04}}.
\end{abstract}

\maketitle

\section{Introduction and Conclusions}

It is undeniable that integrability  of strings propagating on $AdS_3 \times S^3 \times T^4$ \cite{Bogdan} (for reviews see  \cite{rev3,Borsato:2016hud,Seibold:2024qkh}) continues to offer opportunities for surprises, keeping on with the tradition inaugurated by $AdS_5$ and $AdS_4$ \cite{Beisertreview}. The exact $S$-matrix theory is one of the richest available \cite{OhlssonSax:2011ms,seealso3,Borsato:2012ss,Borsato:2013qpa,Borsato:2014hja,Borsato:2013hoa} and the presence of massless modes \cite{Sax:2012jv,Majum,Diego}, see also \cite{Lloyd:2013wza} and \cite{Ben,Ben2}, keeps pushing the technology further and further. The Quantum Spectral Curve (QSC)  \cite{QSC}, see also \cite{Cavaglia:2022xld}, is undergoing an expansion of its methods to accommodate this model \cite{Ekhammar:2024kzp}. Meanwhile \cite{AleSSergey}, see also \cite{Seibold:2022mgg,Polvararel,Ker}, have considerably revitalised the field with new algebraic data and have created an exciting race with significant progress on all sides  \cite{recent,riab,gaber2,frolo,sei2,sei3,free,Bocconcello,Riccardo,exchange,gamma,gamma2}.

A very important feature of this model is the possibility of a deformation in the form of mixed-flux extension \cite{Cagnazzo:2012se,s1,s2,Babichenko:2014yaa,seealso12} which shows up in the  dispersion relation of the elementary scattering particles \cite{ArkadyBenStepanchuk}:
\begin{equation}
\label{eq:disp-rel}
E = \sqrt{\Big(m + \frac{k}{2 \pi} p\Big)^2 + 4 h^2 \sin^2\frac{p}{2}}.
\end{equation}
Notably $k \in \mathbbmss{Z}$ stands for the WZW level and $m$ is for 
the mass of the particle. An important wealth of work on this and similar deformations has now been done \cite{Baggio}. As it is customary, one calls {\it pure RR} the $k=0$ case (and {\it mixed-flux} otherwise). 

There is a special subarea inside this field currently involving a very intense activity, and that is the relativistic sector. This goes from the BMN limit being non-trivial for same-chirality massless scattering \cite{Diego}, where it describes a CFT {\it \'a l\'a} Zamolodchikov \cite{Zamomassless}, to the entire massless sector being of difference form \cite{gamma,gamma2,Majum,AleSSergey}, to the mixed-flux scattering being amenable to a variety of massive relativistic limits \cite{gamma2,Polvararel,Ker}, all pointing toward famous integrable $S$-matrices of the past - with crucial  differences. The observation \cite{Diego} is that the massless $S$-matrix of the pure RR theory is very similar but not quite the same as Fendley-Intriligator's ${\cal{N}}=2$ $S$-matrix \cite{ilmondo,ilmondo2,ilmondo3}. In turn, this ${\cal{N}}=2$ $S$-matrix is related to Sine-Gordon taken at a special coupling \cite{BL}, and it explicitly appeared in the paper by Mussardo \cite{Mussardo}, where  important properties in connection with Sine-Gordon at a special coupling are outlined. Considerations on the connections with the ${\cal{N}}=2$ literature in the case of mixed-flux are also expanded upon in \cite{Polvararel}, and in particular in \cite{Ker}, where the emphasis is on the differences in the global structure of the supersymmetry algebra. 

The focus of this paper is instead the difference in the nature itself of the supersymmetry, in particular the appearance of fractional statistics. We review the detailed reason why supersymmetry emerges but is not exactly that which one has in pure RR $AdS_3$, due to a different assignment of fermionic number of the particles. We use the coproduct of the underlying quantum affine symmetry of Sine-Gordon to show this, following the presentation in Bernard and LeClair \cite{BL}, see also section 2 of \cite{ilmondo} and \cite{Cecotti}. We use this to obtain an interpolating $S$-matrix between massless $AdS_3$ and Sine-Gordon. 

Then we deformed this interpolating $S$-matrix in order to relax the special coupling and let it be generic. The aim was to obtain an interpolation to the (or one of the) mixed flux $S$-matrix (-ces). We managed to find two distinct classes of interpolations which solve the Yang-Baxter equation (YBE) all the way in-between. The first class, denoted by $S_\alpha$, corresponds to deformations of Sine-Gordon at generic $q$, and contains the Fendley-Intriligator case and the massless $AdS_3$ as limiting cases. The second class, denoted by $S_{\alpha+}$ in its most general incarnation, corresponds to deformations of Sine-Gordon at the special coupling $q=-i$, and contains the Fendley-Intriligator case, the massless $AdS_3$ and the relativistic mixed-flux as limiting cases. They are all relativistic invariant, and satisfy phyical unitarity for a choice of parameters. Both classes display fractional statistics for the scattering particles. These $S$-matrices are all regular, namely $S(\theta) \to \mathbbmss{1}\otimes \mathbbmss{1}$ as $\theta \to 0$, , with $\theta = \theta_1 - \theta_2$ being the relativistic rapidity-difference of the two scattering particles. This means that they sit within the classification of \cite{corc}, possibly section 4.3.1 modulo twists \footnote{They are also all of free-fermion type, which is consistent with having all the AdS/CFT $S$-matrix of free-fermion type - and also the Sine-Gordon $S$-matrix at special coupling $\xi \to 2\pi$, since Sine-Gordon satisfies $\mbox{const.} = 4 \cos^2 \frac{\pi^2}{\xi} $ in formula (2.2) of \cite{free}}. They all have  $\mathfrak{sl}_2$-type quantum symmetry, they are braiding unitarity in deference to their quantum group origin,  and they can be rephrased in terms of particles with fractional statistics \footnote{Statistics in $1+1$ dimension is a fascinating subject, see for instance \cite{Fro,exchange}.}. Inspired by \cite{exchange}, we highlight this fact by calling them all {\it fractional} $S$-matrices - what seems to be a ubiquitous feature at this point \cite{ilmondo2}.   
 We believe that this work serves a useful purpose to explicitly embed all the relativistic $S$-matrices appearing from the study of string theory in $AdS_3$ inside one single parent relativistic model, which can actually also exactly include Fendley-Intriligator, a.k.a. Sine-Gordon at a special value, showing the common root in terms of fractional statistics.  
It will be of paramount importance to relate our  interpolations with the ones which are being pursued in the literature from the worldsheet viewpoint, such as \cite{Seibold:2024qkh,Bocconcello,Riccardo,sei2} and the very resembling \cite{Ben2}. This paper treats one particular block, corresponding to the left-moving kinematics for both scattering particles - in Zamolodchikov's picture of massless scattering \cite{Zamomassless}, this is a non-perturbative effect which does not correspond to a perturbative classical configuration. We choose to explicitly display one particular block for definiteness and since we are describing a principle.  Incorporating different blocks and bound states \cite{Polvararel} shall promptly follow.   

It would also be very interesting to ascertain whether there is a natural way to identify the fractional statistics of the excitations within the structure of the QSC as constructed in \cite{QSC,Cavaglia:2022xld,Ekhammar:2024kzp}.

\section{Sine-Gordon at special coupling and $AdS_3$}

Let us begin by demonstrating how the Sine-Gordon model taken at a special value of the coupling develops an ${\cal{N}}=2$ supersymmetry. More precisely, the statement which we will review the proof of is the following: at the special coupling $\beta^2 = \frac{16 \pi}{3}$ the usual quantum group symmetry of the Sine-Gordon model becomes isomorphic to an ${\cal{N}}=2$ supersymmetry in the soliton-antisoliton sector \footnote{As we will shortly review, this is completely internal to Sine-Gordon itself, spontaneously emergent so to say. This is not what is often called {\it ${\cal{N}}=2$ supersymmetric Sine-Gordon} (references can be found for instance in \cite{Diego}), which is instead an extension of Sine-Gordon including more fields to make the Lagrangian supersymmetric to start with. In our case, the Lagrangian is exclusively Sine-Gordon, and the supersymmetry emerges for the special value of $\beta$.}, with fractional fermionic number assignment of the states $|\phi| = \mbox{soliton} = \frac{1}{2}$ and  $|\psi| = \mbox{antisoliton} = -\frac{1}{2}$.  

Let us recall that the soliton-antisoliton Sine-Gordon $S$-matrix \cite{Zamolodchikov:1978xm} (for a review see for instance \cite{book}) takes the form
\begin{eqnarray}
S = \Phi(\theta)\begin{pmatrix}1&0&0&0\\0&S_R&S_T&0\\0&S_T&S_R&0\\0&0&0&1\end{pmatrix},\label{upper}
\end{eqnarray}
where
\begin{eqnarray}
&&S_T = \frac{\sinh \frac{\pi \theta}{\xi}}{\sinh \frac{\pi (i\pi-\theta)}{\xi}} , \qquad S_R = \frac{i \sin \frac{\pi^2}{\xi}}{\sinh \frac{\pi (i\pi-\theta)}{\xi}} , \nonumber\\
&&\qquad \theta = \theta_1 - \theta_2,\label{lower}
\end{eqnarray}
written with the {\it in}-basis being $\{|\phi\rangle \otimes |\phi\rangle,|\phi\rangle \otimes |\psi\rangle,|\psi\rangle \otimes |\phi\rangle,|\psi\rangle \otimes |\psi\rangle$\}. The variable $\theta_i$ is the relativistic rapidity of the $i$-th particle, and $\xi = \frac{\beta^2}{8} \frac{1}{1-\frac{\beta^2}{8 \pi}}$. 

We have made the choice in this article to always call the states
\begin{eqnarray}
|\phi\rangle = \begin{pmatrix}1\\0\end{pmatrix}, \qquad |\psi\rangle = \begin{pmatrix}0\\1\end{pmatrix},
\end{eqnarray} 
irrespective or their fermionic grading in different pictures/theories. The context should hopefully help assigning the correct grading, in fact we will try to explicitly say it every time. This is to unclutter the notation. 

{\small

\bigskip

{\small 

{\bf Proof of fractional  supersymmetry}

\medskip

The soliton-antisoliton Sine-Gordon $S$-matrix has quantum affine symmetry $U_q\big(\widehat{\alg{sl}}(2)\big)$ \cite{BL}. To start, let us focus on one particular generator $\hat{E}$, such that
\begin{eqnarray}
\Delta_{21}\hat{E} \, \,S = S\, \,\Delta \hat{E}, \label{such}
\end{eqnarray}
where $\Delta_{21}$ is the coproduct applied to the permuted combination of particles $2 \otimes 1$, and with
\begin{eqnarray}
\Delta\hat{E}= \hat{E} \otimes \mathbbmss{1} + q^{-2H} \otimes \hat{E}, 
\end{eqnarray}
and where $\hat{E}$ is represented by the matrix $x_i E_{\phi \psi}$, having set $x_i = \exp [\frac{\pi \theta_i}{\xi}]$, for the particle $i=1,2$. $E_{\rho\tau}$ is the matrix with only one non-zero entry equal to $1$ in row $\rho$, column $\tau$. We also have $q = \exp \Big[\frac{8 \pi^2 i}{\beta^2}\Big]$, and $H = \frac{1}{2}(E_{\phi \phi} - E_{\psi \psi})$. We then recall that $\Delta(.)$ is the coproduct, in such a way that (\ref{such}) is our notation for the standard condition of invariance of the $S$-matrix under the two-particle representation of the symmetry \cite{BL}.

At the special supersymmetric value $\beta^2  = \frac{16 \pi}{3}$ we see that $q=-i$. 
The $S$-matrix reduces to
\begin{eqnarray}
S = \Phi(\theta) \begin{pmatrix}1&0&0&0\\0&\mbox{sech} \frac{\theta}{2}&-i\tanh \frac{\theta}{2}&0\\0&-i\tanh \frac{\theta}{2}&\mbox{sech} \frac{\theta}{2}&0\\0&0&0&1\end{pmatrix}.\label{FI0}
\end{eqnarray}
We also have
\begin{eqnarray}
&&\Delta\hat{E} |\phi\rangle \otimes |\psi\rangle =(\hat{E} \otimes \mathbbmss{1} + q^{-2H} \otimes \hat{E}) |\phi\rangle \otimes |\psi\rangle \nonumber\\
&&\qquad \qquad \qquad  =  0+(-1)^{\frac{1}{2}} \, x_2 \, |\phi\rangle \otimes |\phi\rangle,\label{queso}
\end{eqnarray}
Both $|\phi\rangle$ and $|\psi\rangle$ have the same fermionic number $|\phi|=|\psi|=0$ in Sine-Gordon, and therefore $\hat{E}$ has zero fermionic number - because it sends the state $|\psi\rangle$ into the state $|\phi\rangle$  \footnote{The fact that soliton and anti-soliton Bethe eigenstates behave like fermions, even though the bare statistics of the vectors in this original picture is $0$, is due the Sine-Gordon $S$-matrix satisfying $S_{\rho\rho}(0)=-1$, $\rho = \phi,\psi$ \cite{here}.}.

We can now rephrase (\ref{queso}) in a new ({\it fractional}) picture:
\begin{eqnarray}
&&\Delta\hat{E}|\phi\rangle \otimes |\psi\rangle = (\mathbbmss{1} \otimes \hat{E} + \hat{E} \otimes \mathbbmss{1}) |\phi\rangle \otimes |\psi\rangle \nonumber\\
&&\qquad \qquad \qquad =(-1)^{\frac{1}{2}} \, x_2 \, |\phi\rangle \otimes |\phi\rangle + 0.\label{quesa}
\end{eqnarray}
Expressions (\ref{queso}) and (\ref{quesa}) giving the same numerical vector is achieved by this: first, $\hat{E}$ is  represented as $x_i E_{\phi \psi} = x_i \begin{pmatrix}0&1\\0&0\end{pmatrix}$, but has fermionic number $|\hat{E}|=1$; second, $|\phi\rangle$ is represented as $\begin{pmatrix}1\\0\end{pmatrix}$, but has fermionic number $|\phi|=\frac{1}{2}$; third, $|\psi\rangle$ is  represented as $\begin{pmatrix}0\\1\end{pmatrix}$, but has fermionic number $|\psi|=-\frac{1}{2}$. This means that
\begin{eqnarray}
&&\Delta\hat{E} |\phi\rangle \otimes |\psi\rangle = (-1)^{\frac{1}{2}} \, x_2 \, |\phi\rangle \otimes |\phi\rangle  \nonumber\\
&&\qquad \qquad \qquad =(-)^{|\hat{E}||1|} |\phi\rangle \otimes \hat{E}|\psi\rangle,\label{quesad}
\end{eqnarray}   
which gives the same result. We see that we have been able to equivalently rephrase the contribution from $q^{-2H}$ in the original coproduct, now in terms of a fractional fermionic number, carried by a redefined value for $|1|$, that is $|1| = |\phi| = \frac{1}{2}$. 

This exercise can be repeated for a partner supercharge $\hat{F} = x_i\, E_{\psi \phi}$, $\Delta\hat{F}= \hat{F} \otimes \mathbbmss{1} + q^{2H} \otimes \hat{F}$. We shall not need the entire ${\cal{N}}=2$ algebra for our scope, but just half will suffice. The  extension of this argument to all the other generators and all other states follows. This reveals the emergence of supersymmetry at this special value of the coupling: we have managed to rephrase the action of the coproduct of the quantum group generators into an equivalent form involving a supercharge, which consistently acts on states with fermionic number $\pm \frac{1}{2}$ in the (graded) tensor product. The coproduct action has become local - as exemplified by (\ref{quesa}). 

In some sense all that has happened is that the quantum-group deformation $q^{-2H}$, resp. $q^{2H}$,  in the coproduct at this value of the coupling become equal to $(-1)^{|\hat{E}|H}$, resp. $(-1)^{|\hat{F}|H}$, with $H=\begin{pmatrix}\frac{1}{2}&0\\0&-\frac{1}{2}\end{pmatrix}$, which simulate the statistical factors which a fractional supersymmetry would produce when swapping the corresponding generators with the states. We have again opted for using the same symbol for the generators represented by the same matrices, despite the fact that they might have different fermionic grading, again hoping that the distinction will always be clear and to lighten the notation. This way it will always be
\begin{eqnarray}
\hat{E} \propto E_{\phi \psi}, \qquad \hat{F} \propto E_{\psi \phi}.
\end{eqnarray} 

}

\subsection{Relation with massless pure-RR $AdS_3 \times S^3 \times T^4$}
The Sine-Gordon $S$-matrix at the special value of the coupling discussed in the previous section reduces to a known $S$-matrix discussed by Fendley and Intriligator \cite{ilmondo,ilmondo2,ilmondo3}. This $S$-matrix is almost the same as the massless \footnote{For definiteness, we focus on right-moving kinematics.} $AdS_3 \times S^3 \times T^4$, but not quite. It turns out that it differs from it by some factors of $\pm i$ \cite{Diego}, and this is because it has the exact same ${\cal{N}}=2$ supersymmetry which we have just described, but the fermionic assignement of the states is different: in fact in $AdS_3$ the fermionic number is the more traditional $|\phi|=0$, and $|\psi|=-1$ (or, equivalently, $|\psi|= 1$ mod $2$) \footnote{This fermionic number is the one used in \cite{Warner}, which therefore ends up being exactly like massless $AdS_3$ although for an incorrect reason (see comment at the very end of section 3 of  \cite{ilmondo2}).}.  Let us summarise here below the situation - notice that we use everywhere in this paper the symbol $\theta_i$ for the particle's relativistic rapidity, even if in the case of massless pure RR $AdS_3$ we mean the relativistic $\gamma_i$ variable \cite{gamma}. Without any index, $\theta$ always indicates the rapidity-difference.
 
Except where it would cause an obvious confusion, we have made the choice to call all the $S$-matrices $S$ and all the dressing phases $\Phi$ regardless of the different cases and models, to avoid a needless proliferation of symbols.

\begin{itemize}

\item Fendley-Intriligator:
\begin{eqnarray}
S = \Phi(\theta) \begin{pmatrix}1&0&0&0\\0&\mbox{sech} \frac{\theta}{2}&-i\tanh \frac{\theta}{2}&0\\0&-i\tanh \frac{\theta}{2}&\mbox{sech} \frac{\theta}{2}&0\\0&0&0&1\end{pmatrix},\label{FI}
\end{eqnarray}
\begin{eqnarray}
&&\Delta{\hat{E}} = \hat{E}\otimes \mathbbmss{1} + \mathbbmss{1} \otimes \hat{E}, \quad \Delta{\hat{F}} = \hat{F} \otimes \mathbbmss{1} + \mathbbmss{1} \otimes \hat{F}, \nonumber\\
&&\qquad |\phi|=\frac{1}{2}, \quad |\psi| = -\frac{1}{2}, \quad |\hat{E}|=1, \quad \nonumber |\hat{F}|=-1,
\end{eqnarray}
with  $\hat{E} = e^{\frac{\theta_i}{2}} E_{\phi \psi}, \hat{F} = e^{\frac{\theta_i}{2}} E_{\psi \phi}$.

\item Right-right massless $AdS_3 \times S^3 \times T^4$:
\begin{eqnarray}
S = \Phi(\theta) \begin{pmatrix}1&0&0&0\\0&\mbox{sech} \frac{\theta}{2}&\tanh \frac{\theta}{2}&0\\0&-\tanh \frac{\theta}{2}&\mbox{sech} \frac{\theta}{2}&0\\0&0&0&1\end{pmatrix},\label{ads3}
\end{eqnarray}
\begin{eqnarray}
&&\Delta{\hat{E}} = \hat{E}\otimes \mathbbmss{1} + \mathbbmss{1} \otimes \hat{E}, \quad \Delta{\hat{F}} = \hat{F} \otimes \mathbbmss{1} + \mathbbmss{1} \otimes \hat{F}, \nonumber\\
&&\quad |\phi|=0, \qquad |\psi| = -1, \quad |\hat{E}|=1, \quad \nonumber |\hat{F}|=-1,
\end{eqnarray}
and again $\hat{E} = e^{\frac{\theta_i}{2}} E_{\phi \psi}, \hat{F} = e^{\frac{\theta_i}{2}} E_{\psi \phi}$.

\end{itemize}

The difference comes only from the fact that the rule for swapping $x$ and $y$ is $(-1)^{|x||y|}$, and we can see that this produces different results in the two cases.

The first step is now to use this, and rewrite the massless $AdS_3$ S-matrix in a way which is even more reminiscent of Sine-Gordon. This is easily achieved by using again the proof which we went through in the previous subsection, since we can just reinstate the Sine-Gordon notation while keeping track of the fermionic number: 

\begin{itemize}

\item Fendley-Intriligator:
\begin{eqnarray}
S= \Phi(\theta) \begin{pmatrix}1&0&0&0\\0&\mbox{sech} \frac{\theta}{2}&-i\tanh \frac{\theta}{2}&0\\0&-i\tanh \frac{\theta}{2}&\mbox{sech} \frac{\theta}{2}&0\\0&0&0&1\end{pmatrix},
\end{eqnarray}
\begin{eqnarray}
&&\Delta{\hat{E}} = \hat{E}\otimes \mathbbmss{1} + q_0^{-2H} \otimes \hat{E}, \quad \Delta{\hat{F}} = \hat{F} \otimes \mathbbmss{1} + q_0^{2H} \otimes \hat{F}, \nonumber\\
&&\quad |\phi|=|\psi|=0, \quad |\hat{E}|=|\hat{F}|=0, \quad q_0=-i \nonumber.
\end{eqnarray}

\item Right-right massless $AdS_3 \times S^3 \times T^4$:
\begin{eqnarray}
S = \Phi(\theta) \begin{pmatrix}1&0&0&0\\0&\mbox{sech} \frac{\theta}{2}&\tanh \frac{\theta}{2}&0\\0&-\tanh \frac{\theta}{2}&\mbox{sech} \frac{\theta}{2}&0\\0&0&0&1\end{pmatrix},
\end{eqnarray}
\begin{eqnarray}
&&\Delta{\hat{E}} = \hat{E}\otimes \mathbbmss{1} + q_0^{2H-\mathbbmss{1}} \otimes \hat{E}, \quad \Delta{\hat{F}} = \hat{F}\otimes \mathbbmss{1} + q_0^{-2H+\mathbbmss{1}} \otimes \hat{F}, \nonumber\\
&& |\phi|=|\psi|=0, \quad |\hat{E}|=|\hat{F}|=0, \quad q_0=-i \nonumber.
\end{eqnarray}
\end{itemize}

\bigskip

{\small 

{\bf Rewriting ordinary supersymmetry}

\medskip

The last statement can be seen for instance considering this argument: take for instance

\begin{eqnarray}
\Delta\hat{E} = \hat{E} \otimes \mathbbmss{1} + {q_0}^{2H-\mathbbmss{1}} \otimes \hat{E}.
\end{eqnarray}
This means
\begin{eqnarray}
&&\Delta\hat{E} |\alpha\rangle \otimes |\psi\rangle = (\hat{E} \otimes \mathbbmss{1} + q^{2H-\mathbbmss{1}} \otimes \hat{E}) |\alpha\rangle \otimes |\psi\rangle =\nonumber\\
&&\hat{E}|\alpha\rangle \otimes |\psi\rangle+(-1)^{\sigma_3 - \mathbbmss{1}} \, x_2 \, |\alpha\rangle \otimes |\phi\rangle, \qquad \alpha=\phi,\psi.\nonumber
\end{eqnarray}
Both $|\phi\rangle=\begin{pmatrix}1\\0\end{pmatrix}$ and $|\psi\rangle=\begin{pmatrix}0\\1\end{pmatrix}$ have fermionic number $|\phi|=|\psi|=0$, and $\hat{E}$ has zero fermionic number - because it sends the state $|\psi\rangle$ into the state $|\phi\rangle$.
We can now rephrase (\ref{queso}) in this way:
\begin{eqnarray}
&&\Delta(\hat{E}) |\phi\rangle \otimes |\psi\rangle = (\mathbbmss{1} \otimes \hat{E} + \hat{E} \otimes \mathbbmss{1}) |\phi\rangle \otimes |\psi\rangle \nonumber\\
&&\qquad \qquad \qquad = x_2 \, |\phi\rangle \otimes |\phi\rangle +\hat{E}|\phi\rangle \otimes |\psi\rangle, \nonumber\\ &&\Delta(\hat{E}) |\psi\rangle \otimes |\psi\rangle = (\mathbbmss{1} \otimes \hat{E} + \hat{E} \otimes \mathbbmss{1}) |\psi\rangle \otimes |\psi\rangle \nonumber\\
&&\qquad \qquad \qquad = -x_2 \, |\psi\rangle \otimes |\phi\rangle+\hat{E}|\psi\rangle \otimes |\psi\rangle.\nonumber
\end{eqnarray}
First, $\hat{E}$ is  represented as $x_i E_{\phi \psi} = e^{\frac{\theta_i}{2}} \begin{pmatrix}0&1\\0&0\end{pmatrix}$, and has fermionic number $|\hat{E}|=1$; second, $|\phi\rangle$ is represented as $\begin{pmatrix}1\\0\end{pmatrix}$, and has fermionic number $|\phi|=0$; third, $|\psi\rangle$ is  represented as $\begin{pmatrix}0\\1\end{pmatrix}$, and has fermionic number $|\psi|=-1$. The other cases work out analogously. Basically this time the quantum-group deformation $q^{2H-\mathbbmss{1}}$, resp. $q^{-2H+\mathbbmss{1}}$,  in the coproduct at this value of the coupling become equal to $(-1)^{|\hat{E}|\Sigma}$, resp. $(-1)^{|\hat{F}|\Sigma}$, with $\Sigma = \begin{pmatrix}0&0\\0&-1\end{pmatrix}$, which simulates ordinary supersymmetry. 

\subsubsection{Going fractional}

It is convenient to obtain an $S$-matrix which depends on an unspecified fractional fermionic grading in such a way to interpolate between Sine-Gordon and $AdS_3$ when $q=-i$:

Fractional interpolation:
\begin{eqnarray}
&&S_\alpha = \Phi_\alpha(\theta) \begin{pmatrix}1&0&0&0\\0&\frac{e^{\frac{\pi \theta}{\xi}}(q^2-1)}{q^2 - e^{\frac{2\pi \theta}{\xi}}}&\frac{-i q e^{-i \pi \alpha}\big(e^{\frac{2 \pi \theta}{\xi}}-1\big)}{q^2 - e^{\frac{2\pi \theta}{\xi}}}&0\\0&\frac{i q e^{i \pi \alpha}\big(e^{\frac{2 \pi \theta}{\xi}}-1\big)}{q^2 - e^{\frac{2\pi \theta}{\xi}}}&\frac{e^{\frac{\pi \theta}{\xi}}(q^2-1)}{q^2 - e^{\frac{2\pi \theta}{\xi}}}&0\\0&0&0&1\end{pmatrix}=\nonumber\\
&&\Phi_\alpha(\theta) \begin{pmatrix}1&0&0&0\\0&S_R&i e^{-i\pi \alpha}S_T&0\\0&-i e^{i\pi \alpha}S_T&S_R&0\\0&0&0&1\end{pmatrix},\nonumber
\end{eqnarray}
\begin{eqnarray}
&&\Delta{\hat{E}} = \hat{E}\otimes \mathbbmss{1} + e^{- 2 H (\log [q] - \log[-i]) + \mbox{fracto}} \otimes \hat{E}, \nonumber\\
&&\qquad \Delta{\hat{F}} = \hat{F} \otimes \mathbbmss{1} + e^{2 H (\log [q] - \log[-i]) - \mbox{fracto}} \otimes \hat{F}, \nonumber\\
&& |\phi|,|\psi| = 0, \quad |\hat{E}|=0, \quad \nonumber |\hat{F}|=0, \nonumber\\
&&\qquad \qquad \qquad \qquad \mbox{fracto} = \log(-1)\begin{pmatrix}\alpha&0\\0&\alpha-1\end{pmatrix},
\end{eqnarray}
with $\hat{E} = e^{\frac{\pi \theta_i}{\xi}} E_{\phi\psi}$, $\hat{F} = e^{\frac{\pi \theta_i}{\xi}} E_{\psi\phi}$, and where $S_R$, $S_T$ are the Sine-Gordon entries (\ref{lower}). 
One can verify that
\begin{eqnarray}
S_\alpha \to \mbox{Fendley-Intriligator (\ref{FI}) if} \, q \to -i, \, \, \xi \to 2 \pi, \, \, \alpha \to \frac{1}{2}\nonumber
\end{eqnarray}
and
\begin{eqnarray}
S_\alpha \to \mbox{massless $AdS_3$ (\ref{ads3}) if} \, q \to -i, \, \, \xi \to 2 \pi, \, \, \alpha \to 0.\nonumber
\end{eqnarray}
We notice that $S_{\alpha}$ solves the Yang-Baxter equation (YBE) for all $\alpha$, and is braiding-unitary $S_\alpha(\theta) S_\alpha(-\theta) = \mathbbmss{1} \otimes \mathbbmss{1}$ if $\Phi_\alpha(\theta)\Phi_\alpha(-\theta)=1$. The parameter $(\alpha,\alpha-1)$ is effectively the statistics of particles $(\phi,\psi)$, which we are using to interpolate between Sine-Gordon (Fendley-Intriligator) $(\frac{1}{2},-\frac{1}{2})$ and $AdS_3$ $(0,-1)$, as $\alpha$ goes from $\frac{1}{2}$ to $0$. 
$S_\alpha$ is also physical-unitary, {\it i.e.} $S_\alpha$ is a unitary matrix for real rapidities, if $\Phi(\theta)$ is a pure phase for real $\theta$ (and $\alpha$ is real.) 

When $S_\alpha$ is taken for generic $q$ and $\alpha = \frac{1}{2}$, it reduces to the ordinary Sine-Gordon $S$-matrix (\ref{upper}). This means that $S_\alpha$ corresponds to a {\it bona fide} deformation of Sine-Gordon theory for particles with {\it varying} fermionic number - in the sense that one can dial the fermionic number as an independent real parameter. In this case we have also managed to derive the crossing-symmetry equation and solve it. To this purpose, as it is the case in $AdS_3$ literature, a second pair of states $(\bar{\phi},\bar{\psi})$ is introduced, and a new $S$-matrix scattering barred states $(\bar{\phi},\bar{\psi})$ against un-barred states $(\phi,\psi)$. We worked it out completely for $q=-i$, the associated $R$-matrix being
\begin{eqnarray}R_{\tiny\mbox{bar-unbar}}=
    \Phi_{mix}(\theta)\begin{pmatrix}-\tanh \frac{\theta}{2}&0&0&-\mbox{sech}\frac{\theta}{2}\\0&e^{-i \pi \alpha} &0&0\\0&0&-e^{i\pi \alpha}&0\\-\mbox{sech}\frac{\theta}{2}&0&0&\tanh \frac{\theta}{2}\end{pmatrix},\nonumber
\end{eqnarray}
such that the crossing equation is
\begin{eqnarray}
    R_{\alpha}(\theta) (C^{-1}\otimes \mathbbmss{1})R^{\mbox{tr}_1}_{\tiny\mbox{bar-unbar}}(\theta + i \pi)(C\otimes \mathbbmss{1}) =  \mathbbmss{1}\otimes \mathbbmss{1},\nonumber
\end{eqnarray}
where $C = \begin{pmatrix}1&0\\0&i\end{pmatrix}$ is the charge-conjugation matrix in one of its possible versions, and $\mbox{tr}_1$ denotes transposition on the first space of the tensor product. Explicit evaluation of this crossing equation produces a condition for the dressing factors:
\begin{eqnarray}
    \Phi_\alpha(\theta)\Phi_{mix}(\theta + i \pi) = - \tanh\frac{\theta}{2},
\end{eqnarray}
which is independent on $\alpha$. This relation therefore admits as a possible solution the same formula as in one of the early treatments of pure massless $AdS_3$ \cite{Diego}, see also \cite{AleSSergey}:
\begin{eqnarray}
    \Phi_{mix}(\theta) = i\Phi_Z(\theta),\qquad \Phi_\alpha(\theta) = \Phi_Z(\theta)
\end{eqnarray}
\begin{eqnarray}
\label{zamo}
\Phi_Z(\theta) = \prod_{\ell=1}^\infty \frac{\Gamma^2(\ell - \tau) \, \Gamma(\frac{1}{2} + \ell + \tau) \,\Gamma(- \frac{1}{2} + \ell + \tau)}{\Gamma^2(\ell + \tau) \, \Gamma(\frac{1}{2} + \ell - \tau) \,\Gamma(- \frac{1}{2} + \ell - \tau)},
\end{eqnarray}
where  
\begin{eqnarray}
\tau \equiv \frac{\theta}{2 \pi i}.
\end{eqnarray}
A complete analysis of poles and zeroes of this dressing factor, and the complete list of singularities of the $S$-matrix, is contained for example in \cite{sei3}. In particular, no bound-state poles are present in the physical strip of the two-particle $S$-matrix, in accordance with the expectations for massless scattering.

\section{Relaxing the special coupling}

The basic idea is now to let $q$ be generic in this reformulation of the problem, and see where this leads. The hope is that this introduces the mixed flux in the $AdS_3$ case - we mean this purely in terms of functional form of the $S$-matrix, since it is clear that the spectra are different  \cite{Polvararel} and it is not immediate to us to imagine how they might smoothly deform one into another. 
We first recalculate the $S$-matrix with generic $q$. Applying this to Fendley-Intriligator of course returns back the ordinary Sine-Gordon. For $AdS_3$ the result, although potentially interesting {\it per se}, is not immediately connected to mixed-flux scattering: this  
is due to the fact that the mixed-flux relativistic coproduct is twisted in a very particular way \cite{gamma2,Polvararel}. 
But there is  a way to obtain {\it exactly} the mixed flux scattering matrix using this philosophy, by adjusting things in such a way that the deformation of the coproduct is precisely the one observed in the mixed flux case. It works like this. We  take the following route to the pure RR massless $AdS_3$ $S$-matrix:
\begin{eqnarray}
&&S_k =\Phi(\theta) 
\begin{pmatrix}1&0&0&0\\0&e^{\frac{\theta}{2}}\big(1 - e^{i \frac{\pi}{k}}\epsilon(\theta)\big)&\epsilon(\theta)&0\\0&\tau_1&e^{\frac{\theta}{2}}\big(1 - e^{i \frac{\pi}{k}}\epsilon(\theta)\big)&0\\0&0&0&\tau_2\end{pmatrix}, \nonumber\\
&&\tau_1 = e^{i\frac{\pi}{k}}\big(1 - e^\theta + e^{i \frac{\pi}{k}+\theta}\epsilon(\theta)\big), \quad \tau_2 = e^{\theta } - e^{i \frac{\pi}{k}}\big(1+e^\theta\big)\epsilon(\theta)),\nonumber\\
&&\qquad \qquad \qquad \Delta{\hat{E}} =\hat{E}\otimes \mathbbmss{1} + q_0^{2H-\mathbbmss{1}}e^{i \pi (q+i)} \otimes \hat{E}, \qquad \nonumber\\
&&\qquad \qquad \qquad \Delta{\hat{F}} = \hat{F}\otimes \mathbbmss{1} + q_0^{-2H+\mathbbmss{1}}e^{-i \pi (q+i)} \otimes \hat{F}, \nonumber\\
&&\qquad \qquad |\phi|=|\psi|=0, \quad |\hat{E}|=|\hat{F}|=0, \quad q_0=-i,\nonumber\\
&&\qquad \qquad q = \frac{1}{k}-i, \quad \hat{E} = e^{\frac{\theta_i}{2}} E_{\phi\psi}, \quad \hat{F} = e^{\frac{\theta_i}{2}} E_{\psi\phi} , \nonumber
\end{eqnarray}

This is obtained using the symmetry alone, and the function $\epsilon(\theta)$ is an arbitrary freedom of the solution not fixed by the symmetry. Solving the Yang-Baxter equation fixes this freedom to
\begin{eqnarray}
\epsilon(\theta) = \frac{1}{e^{\frac{i \pi}{k}}+\frac{1}{e^g - e^{g + \theta}}}, \qquad g \, \, \mbox{constant}.
\label{cho}
\end{eqnarray}
We have solved YBE by noticing that it consistently reduces to a single condition for the function $\epsilon(\theta)$ with various arguments. That is to say that every entry of the Yang-Baxter equation are in fact all one and the same scalar equation for the single function $\epsilon(\theta)$. The form of this equation is $F\Big(\epsilon(\theta_1-\theta_2),\epsilon(\theta_1-\theta_3),\epsilon(\theta_2-\theta_3),...\Big)=0$, with $F$ a certain scalar algebraic expression (the dots denote the explicit $\theta_i$ and $k$ dependence coming from the other parts of the $S$-matrix). Having ascertained this fact by brute force, the rest is not difficult: essentially  one can for instance solve for $\epsilon(\theta_1-\theta_2)$, and impose the result to be independent of $\theta_3$ (by taking the derivative w.r.t. $\theta_3$ and setting it to zero). This can be repeated for all the variables. In the resulting equations, one can then set in turn each of the $\theta_i$  to zero and permute the arguments. Some simple manipulations ultimately allow one to reduce to an ordinary differential  equation for the function $\epsilon(\theta)$ alone: one can then solve this differential equation, and one obtains the solution (\ref{cho}).

To obtain a limit to the pure RR massless $S$-matrix we can do 
\begin{eqnarray}
    &&\lim_{\substack{k \to \infty \\ \epsilon(\theta) \to \tanh\frac{\theta}{2}}}    S_k \, \, = \, \, \Phi(\theta)\begin{pmatrix}1&0&0&0\\0&\mbox{sech} \frac{\theta}{2}&\tanh \frac{\theta}{2}&0\\0&-\tanh \frac{\theta}{2}&\mbox{sech} \frac{\theta}{2}&0\\0&0&0&1\end{pmatrix}, \label{lim1}\\
&&\Delta{\hat{E}} \to \hat{E}\otimes \mathbbmss{1} + q_0^{2H-\mathbbmss{1}} \otimes \hat{E}, \qquad \Delta{\hat{F}} \to\hat{F}\otimes \mathbbmss{1} + q_0^{-2H+\mathbbmss{1}} \otimes \hat{F}.\nonumber
\end{eqnarray}
The choice displayed in the limit (\ref{lim1}) consistently corresponds to setting $k \to \infty$, and then setting $e^g = -\frac{1}{2}$, in (\ref{cho}).

We can instead refrain from taking any limit on $q$:
\begin{eqnarray}
&&S_k =\Phi(\theta) \begin{pmatrix}1&0&0&0\\0&e^{\frac{\theta}{2}}\big(1 - e^{i \frac{\pi}{k}}\epsilon(\theta)\big)&\epsilon(\theta)&0\\0&\tau_1&e^{\frac{\theta}{2}}\big(1 - e^{i \frac{\pi}{k}}\epsilon(\theta)\big)&0\\0&0&0&\tau_2\end{pmatrix}, \nonumber\\
&&\tau_1 = e^{i\frac{\pi}{k}}\big(1 - e^\theta + e^{i \frac{\pi}{k}+\theta}\epsilon(\theta)\big), \quad \tau_2 = e^{\theta } - e^{i \frac{\pi}{k}}\big(1+e^\theta\big)\epsilon(\theta)),\nonumber\\
&&\qquad \qquad \qquad \Delta{\hat{E}} =\hat{E}\otimes \mathbbmss{1} + q_0^{2H-\mathbbmss{1}}e^{i \pi (q+i)} \otimes \hat{E}, \qquad \nonumber\\
&&\qquad \qquad \qquad \Delta{\hat{F}} = \hat{F}\otimes \mathbbmss{1} + q_0^{-2H+\mathbbmss{1}}e^{-i \pi (q+i)} \otimes \hat{F}, \nonumber\\
&&\qquad \qquad|\phi|=|\psi|=0, \quad |\hat{E}|=|\hat{F}|=0, \quad q_0=-i, \qquad \nonumber\\
&&\qquad \qquad q = \frac{1}{k}-i, \quad \hat{E} = e^{\frac{\theta_i}{2}} E_{\phi\psi}, \quad \hat{F} = e^{\frac{\theta_i}{2}} E_{\psi\phi} , \nonumber
\end{eqnarray}
\begin{eqnarray}
&&\lim\limits_{ \epsilon(\theta) \to - i \frac{\sinh\frac{\theta}{2}}{\sin \big(\frac{\pi}{k} - \frac{i \theta}{2}\big)}} S_k \, \, \,  = \, \, \, \qquad \qquad \qquad \qquad \qquad \qquad \qquad \nonumber\\
&&\Phi(\theta) \begin{pmatrix}1&0&0&0\\0&\frac{i \sin \frac{\pi}{k}}{\sinh \big(\frac{i \pi}{k} + \frac{\theta}{2}\big)}&\frac{\sinh \frac{\theta}{2}}{\sinh \big(\frac{i \pi}{k} + \frac{\theta}{2}\big)}&0\\0&\frac{\sinh \frac{\theta}{2}}{\sinh \big(\frac{i \pi}{k} + \frac{\theta}{2}\big)}&\frac{i \sin \frac{\pi}{k}}{\sinh \big(\frac{i \pi}{k} + \frac{\theta}{2}\big)}&0\\0&0&0&\frac{\sinh \big(\frac{i \pi}{k} - \frac{\theta}{2}\big)}{\sinh \big(\frac{i \pi}{k} + \frac{\theta}{2}\big)}\end{pmatrix}.\label{lim2}
\end{eqnarray}
The limit coincides with the left-left relativistic mixed-flux scattering - see for instance \cite{Polvararel}.
The choice displayed in the limit (\ref{lim2}) consistently corresponds to setting $e^g = \frac{i}{2} \mbox{csc} \frac{\pi}{k}$, in (\ref{cho}). 

To recap, we have an interpolation between pure RR massless and relativistic mixed-flux using the function $\epsilon(\theta)$, which solves the YBE all the way in-between:
\begin{eqnarray}
&&S_k = \Phi(\theta) \times \nonumber\\
&&\begin{pmatrix}1&0&0&0\\0&\frac{e^{\frac{\theta}{2}}}{1 - e^{i \frac{\pi}{k} + g}\big(e^\theta-1\big)}&\frac{1}{e^{\frac{i \pi}{k}}+\frac{1}{e^g - e^{g + \theta}}}&0\\0&\frac{e^{i \frac{\pi}{k}}\big(1-e^\theta\big)\big(1 + e^{i \frac{\pi}{k}+g}\big)}{1 - e^{i \frac{\pi}{k} + g}\big(e^\theta-1\big)}&\frac{e^{\frac{\theta}{2}}}{1 - e^{i \frac{\pi}{k} + g}\big(e^\theta-1\big)}&0\\0&0&0&\frac{e^{\theta} +e^{i \frac{\pi}{k} + g}\big(e^\theta-1\big)}{1 - e^{i \frac{\pi}{k} + g}\big(e^\theta-1\big)}\end{pmatrix},\nonumber
\end{eqnarray}
\centerline{\includegraphics[width=9cm]{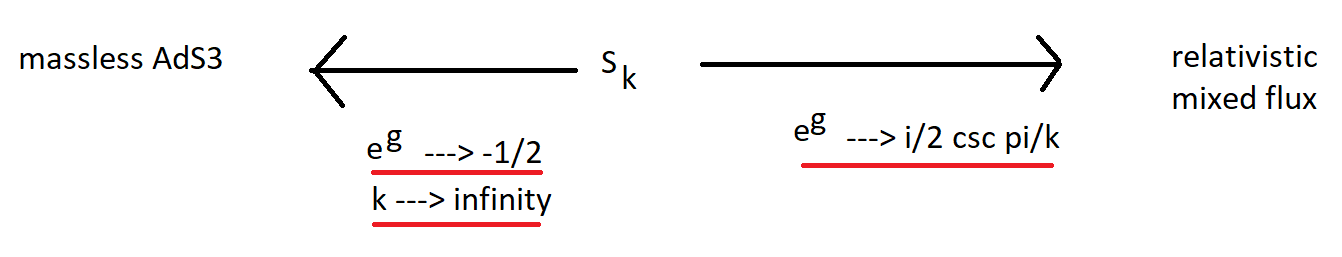}}
 Notice also that $S_k$ satisfies braiding unitarity provided $\Phi(\theta)\Phi(-\theta)=1$. 
It is also physical-unitary if $\Phi(\theta)$ is a pure phase for real $\theta$ and real $k$, and if a certain condition holds on $g$, specifically
\begin{eqnarray}
\mbox{Im}(g) = -i \log \Big[-\frac{1}{2} e^{-\mbox{Re}(g) - \frac{i \pi}{k}}\Big(1+\sqrt{1 - 4 e^{2 \mbox{Re}(g)}}\Big)\Big],\nonumber
\end{eqnarray}
which is real as long as $\mbox{Re}(g) \geq -\log 2$. The latter condition is satisfied (at least) for both the pure RR and the mixed-flux values of $g$.

{\small 

\bigskip

In mathematical terms one can introduce a Drinfeld twist to transform the $R$-matrices, for example
\begin{eqnarray}
&&\hat{E}\otimes \mathbbmss{1} + q_0^{2H-\mathbbmss{1}}e^{i \pi (q+i)} \otimes \hat{E} = T (\hat{E}\otimes \mathbbmss{1} + q_0^{2H-\mathbbmss{1}} \otimes \hat{E})T^{-1}, \nonumber\\
&&\hat{F}\otimes \mathbbmss{1} + q_0^{-2H+\mathbbmss{1}}e^{-i \pi (q+i)} \otimes \hat{F}= T (\hat{F}\otimes \mathbbmss{1} + q_0^{-2H+\mathbbmss{1}} \otimes \hat{F})T^{-1}\nonumber\\
&&\qquad T = e^{i \frac{\pi (q+i)}{2 } \mathbbmss{1} \otimes \sigma_3},\qquad q = \frac{1}{k}-i,
\end{eqnarray}
where $\sigma_3 = E_{11}-E_{22}$. From the general theory of Drinfeld twists one therefore obtains the relation
\begin{eqnarray}
&&R_k = P S_k, \qquad P =  E_{ij}\otimes E_{ji},\qquad 
T_{21}^{-1} R_k T = \bar{R}. 
\end{eqnarray}
Similar considerations are at the basis of \cite{NietoGarcia:2020tma}, see also \cite{Seibold:2024qkh} and section 4.3 of \cite{Riccardo} for example. Notice that one can use either the $S$-matrix or the associated $R$-matrix - we have actually used the appropriate $R$-matrices throughout the paper whenever we needed to prove that the Yang-Baxter equation 
\begin{eqnarray}
&&R_{12}(\theta_1 - \theta_2)R_{13}(\theta_1 - \theta_3)R_{23}(\theta_2 - \theta_3)=\nonumber\\
&&\qquad \qquad R_{23}(\theta_2 - \theta_3)R_{13}(\theta_1 - \theta_3)R_{12}(\theta_1 - \theta_2)
\end{eqnarray}
is satisfied in the tensor product space of three particles. 
The connection with the theory of Drinfeld twists suggests that some of the deformations which we are presenting in this paper are associated with a change in the integrable boundary conditions in the sigma-model picture. One cannot speak of a spin-chain properly in the case of massless scattering \cite{Diego}, but rather one should interpret this as an $S$-matrix encoding a continuum model near criticality \cite{Zamomassless}.
}

\subsection{Fractional route}

It is again interesting to obtain an $S$-matrix which depends on a (more general) unspecified fractional fermionic number in such a way to interpolate between Sine-Gordon and $AdS_3$ when $q=-i$, but then to obtain the mixed flux $S$-matrix in another regime.

More general fractional interpolation:
\begin{eqnarray}
&&S_{\alpha +} = \Phi(\theta) \times \nonumber\\
&&\begin{pmatrix}1&0&0&0\\0&\frac{e^{\frac{i\pi (\alpha + \beta)}{2}}\big(-1+e^{\frac{\theta}{2}}\kappa(\theta)\big)}{\sigma}&\kappa(\theta)&0\\0&\kappa(\theta)&\sigma e^{\frac{-i\pi (\alpha + \beta)}{2}}\big(-1+e^{-\frac{\theta}{2}}\kappa(\theta)\big)&0\\0&0&0&\tau\end{pmatrix},\nonumber\\
&&\tau = -1 + 2 \kappa (\theta)\cosh \frac{\theta}{2},
\end{eqnarray}
subject to
\begin{eqnarray}
\sigma^2 = 1, \qquad 
\end{eqnarray}
\begin{eqnarray}
&&\Delta{\hat{E}} = \hat{E}\otimes \mathbbmss{1} + e^{- 2 H (\log [q] - \log[-i]) + \mbox{Fracto}} \otimes \hat{E}, \qquad \nonumber\\
&&\Delta{\hat{F}} = \hat{F} \otimes \mathbbmss{1} + e^{2 H (\log [q] - \log[-i]) - \mbox{Fracto}} \otimes \hat{F}, \nonumber\\
&& |\phi|,|\psi| = 0, \quad |\hat{E}|=0, \quad \nonumber |\hat{F}|=0, ,\nonumber\\
&&\hat{E} = e^{\frac{\theta_i}{2}} E_{\phi\psi}, \quad \hat{F} = e^{\frac{\theta_i}{2}} E_{\psi\phi}, \, \, \, \, \mbox{Fracto} = \log(-1)\begin{pmatrix}\alpha&0\\0&\beta-1\end{pmatrix},\nonumber
\end{eqnarray}
and subject to the condition 
\begin{eqnarray}
e^{i \pi \alpha} = - q^2 \, e^{i \pi \beta}, \qquad q = i \sigma e^{\frac{i \pi }{2}(\alpha - \beta)}.\label{occhio}
\end{eqnarray}
The function $\kappa(\theta)$ is again an arbitrary freedom of the symmetry solution, restricted by YBE as we will show shortly. 

One can verify that
\begin{eqnarray}
&&S_{\alpha +} \to \mbox{Fendley-Intriligator if} \,  \nonumber\\
&&q \to -i, \, \, \alpha,\beta \to \frac{1}{2}, \quad \kappa(\theta) = \mbox{sech} \frac{\theta}{2}, \qquad \sigma = - 1,
\end{eqnarray}
and
\begin{eqnarray}
&&S_{\alpha +}  \to \mbox{massless $AdS_3$  if} \,  \nonumber\\
&&q \to -i, \, \, \alpha,\beta \to 0,\quad \kappa(\theta) = \mbox{sech} \frac{\theta}{2}, \qquad \sigma = - 1.
\end{eqnarray}
Finally, by choosing
\begin{eqnarray}
\sigma = - e^{\frac{i \pi (\alpha + \beta )}{2} - \frac{i \pi}{k}}, \qquad \kappa(\theta) = \frac{e^{\frac{\theta}{2}}\big(e^{\frac{2 i \pi}{k}}\ - 1\big)}{e^{\frac{2 i \pi}{k} +\theta} - 1},
\end{eqnarray}
we see that 
\begin{eqnarray}
S_{\alpha +} \to \Phi(\theta) \begin{pmatrix}1&0&0&0\\0&\frac{i \sin \frac{\pi}{k}}{\sinh \big(\frac{i \pi}{k} + \frac{\theta}{2}\big)}&\frac{\sinh \frac{\theta}{2}}{\sinh \big(\frac{i \pi}{k} + \frac{\theta}{2}\big)}&0\\0&\frac{\sinh \frac{\theta}{2}}{\sinh \big(\frac{i \pi}{k} + \frac{\theta}{2}\big)}&\frac{i \sin \frac{\pi}{k}}{\sinh \big(\frac{i \pi}{k} + \frac{\theta}{2}\big)}&0\\0&0&0&\frac{\sinh \big(\frac{i \pi}{k} - \frac{\theta}{2}\big)}{\sinh \big(\frac{i \pi}{k} + \frac{\theta}{2}\big)}\end{pmatrix},\nonumber
\end{eqnarray} 
which is exactly the mixed-flux result. Given that $\sigma$ is a sign, we must have
\begin{eqnarray}
&&\beta = \frac{1}{k} - \alpha - \frac{2 i \log (-\sigma)}{\pi} + 4 n  = \nonumber\\
&&\quad (2+\frac{2}{k} - \alpha + 4 n, \frac{2}{k} - \alpha + 4 n), \, \, n \in \mathbbmss{Z}, \, \sigma =(+1,-1).\nonumber
\end{eqnarray}
In this case it is still possible to have $\beta=\alpha$ provided that
\begin{eqnarray}
\alpha = \beta = (\frac{1}{k} + 2n + 1, \frac{1}{k} + 2n),\qquad \qquad n \in \mathbbmss{Z},
\end{eqnarray} 
which all the way confirms that there is a fractional picture where the fractional supersymmetry is either
\begin{eqnarray}
|\phi| = \frac{1}{k} + 1, \qquad |\psi| = 
\frac{1}{k},\label{31}
\end{eqnarray}
or
\begin{eqnarray}
|\phi| = \frac{1}{k}, \qquad |\psi| =\frac{1}{k}-1. \label{32}
\end{eqnarray}
The two choices (\ref{31}) and (\ref{32}) are in fact mathematically isomorphic, being simply related to each other by a $\mathbb{Z}_2$ symmetry which swaps the two basis elements combined with a $2\pi$ argument-shift, and they are therefore physically equivalent. 

The general solution interpolates between Fendley-Intriligator, pure RR and mixed flux $AdS_3$. We impose once again the Yang-Baxter equation to fix the residual freedom in general fashion. By similar techniques as we have used in the previous section we find that the unique solution is
\begin{eqnarray}
\kappa(\theta) = \frac{e^{\frac{\theta}{2}}}{1 + 2 \, d \, e^{\frac{\theta}{2}} \sinh \frac{\theta}{2}}, \qquad d \, \ \mbox{constant}. \label{chu}
\end{eqnarray} 
The interpolation is therefore
\begin{eqnarray}
&&S_{\alpha +} = \Phi(\theta) \times\nonumber \\
&&
\begin{pmatrix}1&0&0&0\\0&\frac{e^{\frac{\theta}{2}}}{1 + d (e^{\theta} -1)}&\frac{(1-d)e^{\frac{i\pi (\alpha + \beta)}{2}}\big(e^{\theta}-1\big)}{\sigma \big(1 + d (e^{\theta} -1)\big)}&0\\0&\frac{d e^{\frac{-i\pi (\alpha + \beta)}{2}}\big(e^{\theta}-1\big)\sigma}{ 1 + d (e^{\theta} -1)}&\frac{e^{\frac{\theta}{2}}}{1 + d (e^{\theta} -1)}&0\\0&0&0&\frac{d(1 - e^{\theta})+e^{\theta}}{1 + d (e^{\theta} -1)}\end{pmatrix},\nonumber
\end{eqnarray}
\centerline{\includegraphics[width=9.5cm]{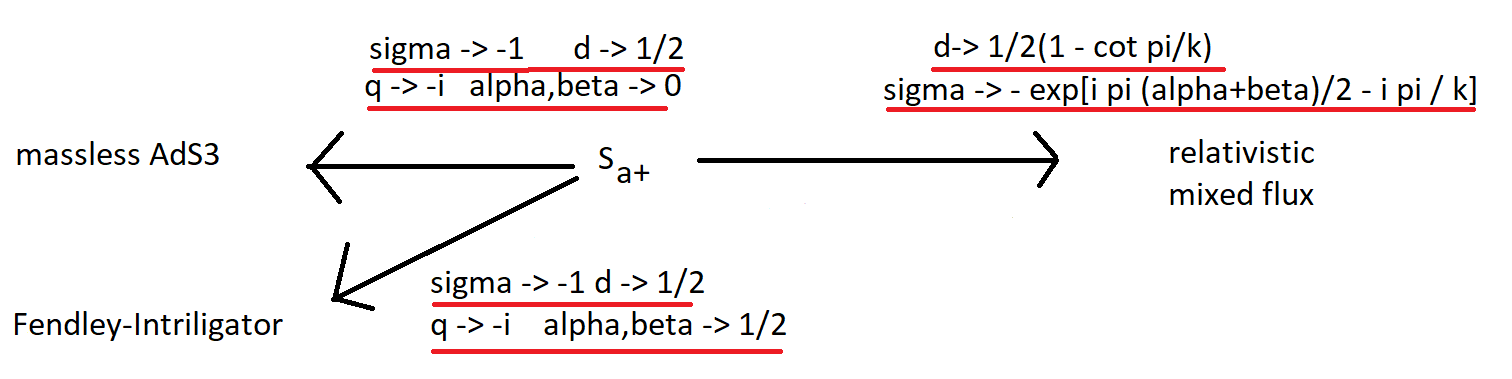}}

 Notice that also $S_{\alpha +}$ satisfies braiding unitarity provided $\Phi(\theta)\Phi(-\theta)=1$.
It is also physical-unitary if $\Phi(\theta)$ is a pure phase for real $\theta$, and if a certain condition holds on $d$, specifically
\begin{eqnarray}
\mbox{Re}(d) = \frac{1}{2},
\end{eqnarray}
\footnote{Unfortunately no million dollars prize is awarded for this.}  and if $\alpha$ and $\beta$ are real.

Finally, it is possible to make a further deformation by replacing everywhere in this section $\theta_i \to \frac{2 \pi}{\xi} \theta_i$, as all the relations continue to hold. 

Therefore in this article we see the introduction  of fractional statistics, a phenomenon characteristic of the physics of lower dimensional devices,  in $AdS_3/CFT_2$ integrable scattering theory. This is taking moves from the relativistic sector where it is recognised in terms of fractional ${\cal{N}}=2$ supersymmetry. 

\section*{Acknowledgments}

We thank Davide Polvara and Bogdan Stefa\'nski, jr., for very interesting discussions. We thank Bogdan Stefa\'nski, jr., for the wisdom and guidance through the Landau-Ginzburg land. We thank the organisers of IGST 2025 at King's for a wonderful conference. We thank Juan Miguel Nieto Garcia for useful correspondence. We thank the anonymous referees for very useful remarks.

No additional research data beyond the data presented and cited in this work
are needed to validate the research findings in this work.

For the purpose of open access, the authors have applied a Creative
Commons Attribution (CC-BY) licence to any author-accepted manuscript version arising.

\end{document}